# HCMUS-Juniors 2020 at Medico Task in MediaEval 2020: Refined Deep Neural Network and U-Net for Polyps Segmentation

Quoc-Huy Trinh[1,3], Minh-Van Nguyen[1,3], Thiet-Gia Huynh[1,3], Minh-Triet Tran[1,2,3]

[1]Faculty of Information Technology, University of Science, VNU-HCM
[2]John von Neumann Institute, VNU-HCM
[3]Vietnam National University, Ho Chi Minh city, Vietnam
{20120013,20127094,20120070}@student.hcmus.edu.vn,tmtriet@fit.hcmus.edu.vn

## ABSTRACT
The Medico: Multimedia Task 2020 focuses on developing an efficient and accurate computer-aided diagnosis system for automatic segmentation [3]. We participate in task 1, polyps segmentation task, which is to develop algorithms for segmenting polyps on a comprehensive dataset. In this task, we propose methods combining Residual module, Inception module, Adaptive Convolutional neural network with U-Net model and PraNet for semantic segmentation of various types of polyps in endoscopic images. We select 5 runs with different architecture and parameters in our methods. Our methods show potential results in accuracy and efficiency through multiple experiments, and our team is in Top 4 best results with Jaccard index of 0.765.

## 1 INTRODUCTION
The goal of *Medico automatic polyp segmentation challenge* is to evaluate various methods for automatic polyp segmentation that can be used to detect and mask out various types of polyps (including irregular, small or flat polyps) with high accuracy.

In this challenge, our goals are to segment the mask of all types of polyps in the dataset. We consider five solutions, corresponding to our five submitted runs. In our first approach, we adopt a simple U-Net architecture [9] to parse masks of polyps. Second, we replace the regular ReLU with Leaky ReLU to deal with dead neurons.[11] Third, to further boost the result, we design an Inception module to extract better features. [10] Fourth, we use a pretrained model with the Resnet50 backbone to build ResUNet, yielding better obtained results. Last, we employ PraNet, a parallel reverse attention network [1] for polyp segmentation in colonoscopy images.

## 2 METHODS
We submit five runs for this task. In each run, we modify and improve different architectures to enhance accuracy and segmentation speed. Initially, we choose U-Net to create the base model that we can test the baseline of the test set. However, the evaluation get low, we decide to develop the U-Net by using Leaky ReLU, but the result gets lower. Next, we replace the simple Convolution block with a more complicated block, the Inception block and Residual block, to improve the accuracy. However, the results are not higher. So, we decide to choose PraNet and get the results as our expectation.

### 2.1 U-Net Architecture
In the first run, to have the overview of the segmentation problem in this task, we use U-Net, which is used widely in medical image segmentation [5] [8], to segment polyps masks and modify it to get a better baseline result. Our overall architecture, which is uncomplicated as the standard U-Net, consists of convolutional encoding and decoding units that take an image as input and produce the segmentation feature maps with respective pixel classes. [8] In our model, we keep the encoding step - which includes 5x5 and 3x3 convolutional blocks and a 1x1 convolutional block with ReLU. However, in this run, to fit the model, we use only one Pooling layer with a small learning rate and batch size.

### 2.2 Leaky ReLU
In the second run, to enhance the result with U-Net, we breakdown and modify the ReLU layer. During the training step, ReLU can cause something known as dead neurons. To overcome this critical defect of dying neurons, we substitute ReLU with another activation function, the Leaky ReLU. We replace all ReLU activation functions with our custom leaky ReLU blocks while preserving the remaining convolutional blocks and layers. Moreover, after each convolutional block, we also add a Leaky ReLU to evaluate extracted features.

### 2.3 ResUNet
In the third run, we propose to extract more features to improve the model. We add a more standard convolutional block to this structure, but it easily to be a vanishing gradient. To achieve consistent training as the depth of the network increases and also prevent those issues, we replace the building blocks of the U-Net architecture with modified residual blocks of convolutional layers.

[2] So our solution is using ResUNet, to prevent the issue and help our training process better. ResUNet architecture combines the strengths of residual units and features concatenate, which help to ease the training of networks and facilitate information propaga-

tion. Dilation convolution is a powerful tool that can enlarge the receptive field of feature points without reducing the resolution of the feature maps. [7]

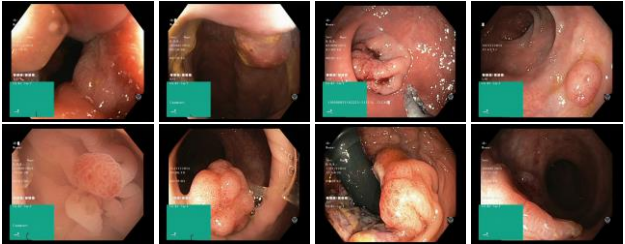

Figure 1: Examples polyps from the test images.

### 2.4 Inception Module

We combine the Inception module with U-Net. The main idea of the Inception architecture based on finding out how an optimal local sparse structure in a convolutional vision network can be approximated and covered by readily available dense components.

One of the main beneficial aspects of this architecture is that it allows for increasing the number of units at each stage significantly without an uncontrolled blow-up in computational complexity. The ubiquitous use of dimension reduction allows for shielding a large number of input filters of the last step to the next layer, first reducing their dimension before convolving over them with large patch size.

Another practically useful aspect of this design is that it aligns with the intuition that visual information should be processed at various scales and then aggregated so that the next stage can abstract features from different scales simultaneously. The efficient allocation of computational resources allows both the width of each step and the total number of steps to be increased without getting computational difficulty. [10] So, we add Inception module to our model. However, to have the compatibility with our U-Net architecture, we modify the order of convolution block to inception module and increase more convolution block to extract feature with a reasonable number to avoid vanishing gradient.

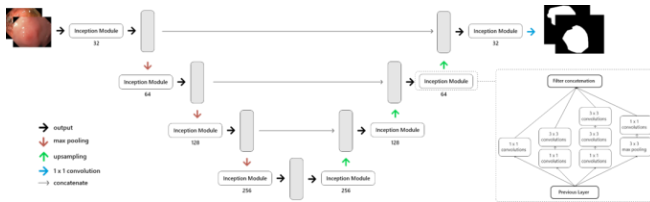

### 2.5 PraNet

In the last run, we use PraNet, which is a parallel reverse attention network for accurate polyp segmentation in colonoscopy and also has the highest benchmark on Medical Image Segmentation on Kvasir-SEG [1] to compare the result to our working.

## 3 DATASET AND EXPERIMENTAL RESULTS

### 3.1 Datasets and Processing

The Kvaris-SEG [4] is a dataset with 1000 images of Polyps in endoscopic images with their mask. This dataset can be downloaded from https://datasets.simula.no/kvasir-seg/.

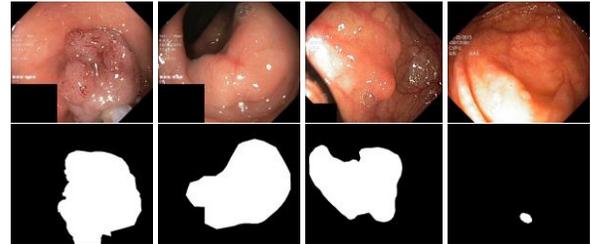

Data preparation: We use Retina net to create a bounding box cover the Polyps in endoscopic images.[6] After generating bounding box, we check the result by our own and create a tool to crop the Polyps from the dataset of endoscopic images and mask in the dataset, this preprocess method help us to increase the quality of images double but keep the resolution of the training and validation set.

Data Augmentation: To increase the numbers of data to improve the training process, we use data augmentation with two methods: rotation and zooming.

Figure 2: Polyps and corresponding masks from Kvasir-SEG.

### 3.2 Results

Evaluation is on the test set of Mediaeval- Medico task, which include 160 images of Polyps in endoscopic images. Table 1 shows that Run 1 is better than Run 2 and Run 3 is better than Run 4. Run 5 is the best model overall. By replace ReLU with Leaky ReLU, although Accuray in Run 2 is not better than Run 1, Recall is gradually higher. Moreover, in 5 runs, ResUNet obtains the result of 90.1, which is approximate with the highest overall is 94.6.

| Run ID | Jaccard | DSC | Recall | Precision | Accuracy | F2 |
|---|---|---|---|---|---|---|
| Run 1 | 0.323 | 0.434 | 0.553 | 0.408 | 0.862 | 0.483 |
| Run 2 | 0.290 | 0.411 | 0.765 | 0.330 | 0.739 | 0.525 |
| Run 3 | 0.406 | 0.515 | 0.507 | 0.757 | 0.901 | 0.501 |
| Run 4 | 0.294 | 0.419 | 0.764 | 0.341 | 0.755 | 0.535 |
| Run5 | 0.766 | 0.841 | 0.894 | 0.844 | 0.946 | 0.857 |

Table 1: MediaEval 2020 challenge's result.

## 4  DISCUSSION

When we use PraNet to test on Medical Images, it performs better than on the camouflage experiment. While PraNet gets high accuracy, but the architecture requires a well-qualified setting to run the model.

## 5  CONCLUSION

Currently, there is a growing interest in the development of computeraided diagnosis (CADx) systems that could act as a second observer and digital assistant for the endoscopists. Algorithmic benchmarking is an efficient approach to analyze. We can try more new method with a different approach and improve the data preparation process to get a higher result.

In general, PraNet still gets the highest accuracy score. However, The computing speed remains high while U-Net and ResUNet take more advantages in speed with slightly different in results. Moreover, data preparation is the most impact process to achieve higher accuracy and evaluation metrics. Therefore, we propose to use Inception Resnet or Densenet 169 feature extractor combine with U-Net to improve the accuracy of the test.

## ACKNOWLEDGMENT

This research is funded by Vietnam National University HoChiMinh City (VNU-HCM) under grant number DS2020-42-01 on "Artificial Intelligence and Extended Reality for Medical Diagnosis and Treatment Assistance".

## REFERENCES

[1] Deng-Ping Fan, Ge-Peng Ji, Tao Zhou, Geng Chen, Huazhu Fu, Jianbing Shen, and Ling Shao. 2020. PraNet: Parallel Reverse Attention Network for Polyp Segmentation. arXiv:2006.11392 [eess.IV]

[2] Zhang X. Ren S. Sun J. He, K. 2016. Identity mappings in deep residual networks. *arXiv preprint arXiv:1603.05027* (2016).

[3] Debesh Jha, Steven A. Hicks, Krister Emanuelsen, Håvard Johansen, Dag Johansen, Thomas de Lange, Michael A. Riegler, and Pål Halvorsen. 2020. Medico Multimedia Task at MediaEval 2020: Automatic Polyp Segmentation. In *Proc. of the MediaEval 2020 Workshop*.

[4] Debesh Jha, Pia H Smedsrud, Michael A Riegler, Pål Halvorsen, Thomas de Lange, Dag Johansen, and Håvard D Johansen. 2020. Kvasir-SEG: A segmented polyp dataset. In *Proc. of International Conference on Multimedia Modeling*. 451–462.

[5] Debesh Jha, Pia H Smedsrud, Michael A Riegler, Dag Johansen, Thomas De Lange, Pål Halvorsen, and Håvard D Johansen. 2019. ResUNet++: An Advanced Architecture for Medical Image Segmentation. In *Proc. of International Symposium on Multimedia*. 225–230.

[6] Tsung-Yi Lin, Priya Goyal, Ross Girshick, Kaiming He, and Piotr Dollár. 2018. Focal Loss for Dense Object Detection. arXiv:1708.02002 [cs.CV]

[7] Z. Liu, R. Feng, L. Wang, Y. Zhong, and L. Cao. 2019. D-Resunet: Resunet and Dilated Convolution for High Resolution Satellite Imagery Road Extraction. 3927–3930.

[8] Chris Yakopcic Tarek M. Taha Md Zahangir Alom, Mahmudul Hasan and Vijayan K. Asari. 2018. Recurrent Residual Convolutional Neural Network based on U-Net (R2U-Net) for Medical Image Segmentation. *arXiv preprint arXiv:1802.06955* (2018).

[9] P. Fischer O. Ronneberger and T. Brox. 2015. U-Net: Convolutional networks for biomedical image segmentation. In *International Conference on Medical Image Computing and Computer-Assisted Intervention*. 234–241.

[10] Christian Szegedy, Wei Liu, Yangqing Jia, Pierre Sermanet, Scott Reed, Dragomir Anguelov, Dumitru Erhan, Vincent Vanhoucke, and Andrew Rabinovich. 2014. Going Deeper with Convolutions. arXiv:1409.4842 [cs.CV]

[11] X. Zhang, Y. Zou, and W. Shi. 2017. Dilated convolution neural network with LeakyReLU for environmental sound classification. In *2017 22nd International Conference on Digital Signal Processing (DSP)*. 1–5.